# Multiple charge beam dynamics in Alternate Phase Focusing structure

S. Dechoudhury[1*], Alok Chakrabarti[1], Y -C.Chao[2]

[1]*Variable Energy Cyclotron Centre, 1/AF Bidhan Nagar, Kolkata 700 064, India*

[2]*TRIUMF, 4004 Wesbrook Mall, Vancouver B.C V6T 2A3, Canada*

*Abstract*

Asymmetrical Alternate Phase (A-APF) focusing realized in a sequence of 36 Superconducting Quarter Wave Resonators has been shown to accelerate almost 81 % of input Uranium beam before foil stripper to an energy of 6.2 MeV/u from 1.3 MeV/u. Ten charge states from *34+* to *43+* could be simultaneously accelerated with the phase of resonators tuned for *34+*. A-APF structure showed unique nature of large potential bucket for charge states higher than that of tuned one. Steering inherent to QWRs can be mitigated by selecting appropriate phase variation of the APF periods and optimization of solenoid field strengths placed in each of the periods. This mitigation facilitates multiple charge state acceleration scheme.

PACS numbers: 29.27. –a,  29.27. -Bd

I  INTRODUCTION

Superconducting quarter wave resonators (QWRs) are widely used for the acceleration of stable as well as rare heavy-ion beams. In ISOL type Rare Ion Beam (RIB) facilities employing linear accelerators for RIB acceleration, superconducting QWR cavities are often used to accelerate the beam to energies enough to carry out Coulomb barrier physics, usually in the range 5 to 7 MeV/u, after initial acceleration in RFQ and room temperature linacs up to energy of about 1 MeV/u. TRIUMF's ISAC II facility [1] is a typical example where this acceleration scheme has been implemented successfully. For RIB acceleration, the transmission efficiency of the accelerating structure is one of the most important considerations since one cannot afford to lose the beam intensity in the process of acceleration. Superconducting QWR (SC QWR) cavities with independent phase and high field levels can accept heavy-ion beams with appreciable transverse and longitudinal emittances and can thus accelerate input beams coming out of the room temperature accelerators of the preceding acceleration stages with practically no loss in the beam intensity. However, to optimally use the high accelerating field of SC QWRs a charge

*sdc@vecc.gov.in*

stripper is used upstream of the SC QWR accelerators to increase the charge state of the unstable heavy-ion beam. This scheme is equally suitable for acceleration of high current Uranium beams to very high energies since it is difficult to get enough intensity for high charge states from the ion source, for example from an ECR Ion source. It is often advantageous to extract comparatively low charge state and higher intensity Uranium beam from the ion-source, accelerate it to about 1 MeV/u and increase the charge state by stripping before accelerating further in SC QWRs. The stripper, however, has an undesirable feature. It produces a number of charge states [2] and one chooses the most abundant one for further acceleration through the QWRs. A major fraction of the beam is thus lost in the process. One can gain manifold in beam intensity if multiple charge states can be simultaneously accelerated. Numerical simulation of a driver Linac for rare ion beam facility accelerating multiple charge states of Uranium beam after stripper has been carried out earlier [3]. Simulation showed that five charge states centered around $q = 75+$ can be simultaneously accelerated from 12.3 MeV/u to 85.5 MeV/u. Also, eight charge states of Uranium beam had been simultaneously accelerated from 286 MeV (1.2 MeV/u) to 690 MeV (2.9 MeV/u) through a section of ATLAS [4]. In this case simultaneous acceleration of multiple charge states has been achieved through comparatively low acceleration gradient; about 400 MeV total energy gain in 24 split co-axial SC resonators.

In the case considered in reference 3, the stripping was considered at higher energy producing an equilibrium charge state of $q = 75 +$ for Uranium. This has an advantage. Owing to high charge to mass ratio the phase offset for other charge states is small when resonators had been tuned for $q = 75 +$. Also the width of charge state distribution is comparatively small at higher energy (~ 12 MeV/u). The situation is much more complicated for stripping at lower energy (say 1.3 MeV/u), since stripping results in a much broader charge distribution around the equilibrium charge state. As an example, Uranium beam stripped at 1.3 MeV/u is distributed over 12 charge states from 32+ to 43+ around the equilibrium charge of 37+. In such cases, simultaneous acceleration of multiple charge states through a long chain of independently phased resonators without compromising the accelerating gradient becomes challenging and difficult. *Asymmetrical Alternate Phase Focusing (A-APF) QWR structure [5, 6, 7],* owing to its inherent larger longitudinal and transverse acceptances seems to be the only candidate that can in principle provide a solution for multiple charge acceleration. It has been shown earlier that by suitably changing the sign of the synchronous phase longitudinal and transverse beam focusing can be realized over a long chain of QWRs [7]. Detailed theoretical analysis corroborated by particle tracking code showed [7] that A-APF scheme can provide appreciable longitudinal and transverse acceptance area and this design has been proposed for the ANURIB facility at VECC [8].

Acceptance of multiple charge acceleration depends on the effective focusing system which can limit the transverse emittance growth [4]. Also, the longitudinal dynamics becomes extremely challenging for simultaneous multi-charge acceleration if one wants to keep a good enough acceleration gradient. A-APF as brought out in our earlier work [7], provides an effective transverse and longitudinal focusing without the need for separate longitudinal (re-buncher) and transverse (solenoids) focusing device even without compromising the acceleration gradient. In case of Quarter Wave Resonators (QWR) steering effect due to the asymmetric field profile would lead to effective emittance growth. It has already been shown [7] that steering effects gets mitigated over a period due to the oscillating phase profile of the A-APF configuration thus limiting the transverse emittance growth facilitating the possibility of multiple charge acceleration. This prompted us to look into the efficacy of A-APF systems in multiple charge acceleration especially with QWRs as the accelerating cavity.

The present paper addresses the design of a multiple charge acceleration scheme through such an A-APF structure realized in a sequence of 36 quarter wave resonators. Changes have been incorporated on the base line design of the earlier A-APF scheme [7] in order to facilitate the option of multiple charge state acceleration. Steering kicks that depend on the energy of the beam and phase of resonator [9] have been found to play a dominant role in case of simultaneous acceleration of multiple charge states. Accordingly, optimization of phase and solenoid fields have been carried out in each cryo- module, which could ensure a common transverse and longitudinal acceptance for most of the post-stripper charge states. The final acceleration scheme consisting of 36 QWRs with $\beta_d$ of 0.06 and 0.1 is shown in *Figure 1* Using this set of QWRs operated in A-APF mode, ten charge states (from *34+* to *43+*) of uranium beam could be accelerated efficiently from 1.3 MeV/u to 6.2 MeV/u (~ 1160 MeV total energy gain in 36 resonators).

## II  DESIGN OF A-APF CONFIGURATION

*Potential bucket for different charge states*

The designed beta ($\beta_d$) for QWR's is selected on the basis of transit time factor over the energy range. We have chosen frequency to be 100 MHz, aperture diameter 20 mm and gap to $\beta_d\lambda$ ratio as 0.2. The choice of designed beta for the resonators ($\beta_d$) as well as the number of resonators in a period takes care of the fact that transit time factor (TTF) remains greater than 0.8 for the entire energy range. In a particular period, the set of resonators have synchronous phase obeying a stepwise function. Using stability analysis the phases and electric field of the resonators in APF periods have been chosen so that cosine of transverse and longitudinal phase advance lies close to 0. This ensured a strong longitudinal and transverse focusing in each such APF periods. Detailed analysis to ascertain these choices have already been covered in reference [7]. The present design exploits this particular feature of strong focusing in both dimensions for achieving multiple charge state acceleration.

The phase acceptance and energy width of any particular charge state can be derived from an effective potential [10]. The effective potential in the present work has been calculated using smooth approximation formalism with acceleration [11]. The on-axis electric fields for these calculations [7] have been simulated using CST [12]. In order to analyze the efficacy of smooth approximation with acceleration in predicting the phase acceptance of the periods, we had carried out particle tracking with real 3D fields of the cavities. Retracing back the successful particles the phase acceptance was determined. The phase acceptance values were found to be within ± 5% of the value predicted by the smooth approximation formalism [7]. The resonator phases have been optimized as per the A-APF scheme for a particular charge state to be termed as the 'tuned' charge state. Once the potential bucket has been calculated for the tuned charge state, one can derive the buckets for other charge states with same phase tuning using the velocity and time profile for other charge states.

The calculations show that the potential bucket becomes shallower (decrease in energy width acceptance) for charge states lower than the tuned one. On the other hand for higher charge states appreciable energy and phase width still exist. So, one can in principle accelerate charge states higher than that of tuned one. The potential buckets for different charge states of Uranium for the first focusing period comprising of 6 QWRs are shown in *Figure 2* when the resonators are tuned for $q = 34+$. It is evident that for charge states higher than $q = 34+$, the depth of the effective potential increases with the charge state. However, there is a flip side - there occur a shift in synchronous phase (for which energy width acceptance is maximum) for higher charge states. This effect is not appreciable for a single period (or a number of resonators) but becomes significant with increase in the number of resonators, decreasing thereby the common acceptance area amongst different charge states. The choice of tuned charge state less than 34+ would lead to inefficient acceleration of dominantly produced charge states after the stripper.

*Steering effect and optimization of solenoids*

In quarter wave resonators (QWR) owing to its resonant structure, the electric and magnetic fields in the accelerating gaps are not axially symmetric. QWRs have in addition to accelerating field (along z), a vertical electric field (Ey) and a horizontal magnetic field (Bx). In case of superconducting cavities, due to high acceleration gradients, both the fields have prominent steering effects on the beam. Steering depends on resonator's synchronous phase, beam velocity and acceleration gradient [9]. The problem of steering becomes more dominant in case of multiple charge state acceleration, as different charge states have different velocity profiles. It reduces the acceptance area and shifts the longitudinal acceptance phase space from that of the tuned one.

Steering being dependent on sinusoid of phases, it changes sign with the sign of individual resonator phases. In a particular APF period as sign of phase oscillates from positive to negative the steering kick gets cancelled [7]. Similarly, if the last QWR of an APF period and first QWR of immediate next period have the same phase sign, steering kick gets amplified in same direction, reducing the common acceptance area for multiple charge state acceleration. Accordingly one needs to change the A-APF scheme that has been studied in our earlier work [7] to facilitate the option of multiple charge state acceleration. We have considered alternating positive and negative cycle of variation in phase, when going from one period to the next as shown in *Figure 3*. This reduces the steering effect to a larger extent. The selected electric field gradient (product of Electric field and TTF) for the cavities in five cryo modules are 5.3 MV/m, 5.2 MV/m, 4.1 MV/m, 3.5 MV/m and 3.8 MV/m respectively.

The *Y* direction steering can be calculated using the electric and magnetic field profile as seen by the particle while traversing through the resonators. The field integration formula for steering described in reference [13] has been utilized to calculate the steering for different charge states. Solenoids placed in each cryo module produce *X - Y* coupling as well as focusing of the beam. Steering caused by QWRs before the solenoids would shift the particle to off-axis trajectory and after solenoid it would enter the next set of QWRs with the combined effect of solenoids and steering. It has been found that X-Y coupling results in transferring a part of Y direction kick to X. Thus solenoids could be tuned within an optimum range so that different on-axis charge states can be close to the axis and have lesser divergence as it exits the period. Coupled equation of motion for solenoids has been described in reference [14]. Using MATHEMATICA [15], equations of motion has been solved for different charge states to find out the transverse co-ordinates and divergence at the end of each period taking into the effect of solenoid focusing and steering in the cavities. *Figure 4* depicts variation of co-ordinate and divergence at the end of second focusing period for q =37+ and width of distribution in transverse co-ordinates for different charge states.

The solenoids in each focusing periods have been optimized so that most dominant charge state *37+* on axis particle undergoes minimum *Y* steering at the end of each APF period. It was also ensured that different charge states (having different velocity profiles through resonators) lie within the close proximity of the dominant charge state at the end of the periods. Such optimization also ensured minimum transverse emittance growth for all charge states combined as compared to single charge acceleration. Solenoid considered here are of length 0.25 m and for maximum optimized field strength of 9 T, field at the nearest QWR (end of preceding QWR and start of succeeding QWR) is only around 120 Gauss which is much below the critical magnetic field of Nb.

## II   BEAM DYNAMICS SIMULATION

*Particle tracking simulation*

The particle tracking have been carried out using GPT [16] with 3D fields of two types ($\beta_d$ = 0.06 and 0.1) of QWRs obtained from CST simulation. Space charge is not considered since for applications considered here the beam intensity is never high enough for the space charge to play a significant role. An aperture of 10 mm has been considered in GPT for particle tracking through all the resonators and solenoids. Optimization algorithm as described in the previous section dictated the values of five solenoids. Beam having distribution in Y-Y$'$ plane and $\Delta E$-$\Delta \phi$ with X = X$'$ = 0 has been tracked for charge states from *34+* to *42+* in order to find common longitudinal acceptance . Gaussian distribution with normalized longitudinal emittance of *4π keV/u - nsec* (phase rms width ~ ± 2deg and energy rms width ~ ± .01 MeV/u) with the orientation same as that of acceptance ellipse, and uniform distribution of transverse emittance (both in X & Y)  *0.2 π mm mrad* have been generated using MATHEMATICA, as the input beam. These values are nearly equal to those measured for ISAC I output beam at TRIUMF [17].

*Multiple charge state acceleration*

Separate particle tracking for each charge states from *34+* to *42+* has been carried out with the set of input particle distribution created. With distribution in *X & X$'$* fine tuning of solenoids close to the optimized values have been done to maximize the transmission efficiency for the dominant charge state *q* = 37+, by an iterative method. Optimized solenoid field in each cryomodule are 6.8 T, 6 T, 7.5 T, 6.0 T and 9 T respectively.

For q = 34+, only 47% are accelerated to energy more than 6 MeV/u, while for higher charge states almost all particles reaching at the end are accelerated to energy in excess of 6 MeV/u. Detailed curve showing the transmission efficiency along with the charge state fraction of all the charges created after stripping is shown in *Figure* 5.

Considering fraction with which a particular charge state is created after stripping at 1.3 MeV/u and fraction of it accelerated to energy greater than 6.0 MeV/u, 81.5 % of the input uranium beam before foil stripper has been finally accelerated to 6.23 MeV/u with FWHM of ± 1.5%. In this configuration 73% of q = 43+ beam can also be accelerated to 6.29 MeV/u with FWHM of ± 0.5% although its charge state fraction is less than 1%. Final transverse and longitudinal distribution of all the charge states at 0.3 m after the last QWR of the fifth period is shown in *Figure* 6. These represent only the fraction that has been accelerated to energy of 6 MeV/u or more. Emittance values along with mean energy of different charge states have been tabulated in Table 1. The transmission efficiency corresponding to different mismatched longitudinal input beam (*Figure* 7) shows an appreciable tolerance factor of  ~ 20% in input beam. Misalignment of resonators and solenoids would also have an adverse effect on transverse beam dynamics and hence on beam transmission efficiency. GPT simulation using 5000 particles distributed equally amongst ten charge states from 34+ to 43+ shows that simultaneous misalignment of all the 36 resonators and five solenoids up to +/-100 μm both in X and Y does not have any appreciable effect on transverse emittance and transmission efficiency. Final transverse emittance for particles finally accelerated to energy greater than 6 MeV/u in X & Y for misaligned sets is shown in *Figure 8*.

IV   CONCLUSION

It is shown that a A-APF structure can be used to accelerate with high efficiency multiple charge states simultaneously through a long chain of Quarter Wave resonators (QWRs) maintaining a high enough acceleration gradient. APF structure showed unique nature of large potential bucket for charge states higher than that of tuned one resulting in good enough longitudinal acceptance for different charge states. Steering inherent to QWRs limits the longitudinal and transverse acceptance of all charge states posing a serious problem. However this effect can be mitigated by selecting appropriate phase variation of APF periods and optimizing solenoids placed in each of the periods. In this design, one solenoid per cryomodule has been found to be enough to transmit the beam with 81% efficiency while it got accelerated from 1.3 MeV/u to ~ 6.2 MeV/u. Increasing the number of solenoids may help marginally to increase the efficiency but this would increase the complexity of the cryo-module and of-course the cost. Present study showed A-APF as a viable and potential candidate for such multiple charge state acceleration starting from charge stripping at low energy ~ 1 MeV/u (inducing larger charge state distribution width) and accelerating to such higher energy ~ 6.2 MeV/u without compromising acceleration gradient.

**Acknowledgement**

This work is a part of the research project funded by the Department of Atomic Energy for the development of ANURIB Facility.

.

# LIST OF TABLES

Table 1. Normalised emittance and energy values for different charge states

| q | Norm. RMS EmitX ($\pi$ mm mrad) | Norm. RMS EmitY ($\pi$ mm mrad) | Norm. RMS EmitZ ($\pi$ keV/u nsec) | Mean Energy (MeV/u) | $\sigma_E$ ($\frac{MeV}{u}$) |
|---|---|---|---|---|---|
| 34+ | 0.1874 | 0.18966 | 0.00383 | 6.37 | 0.055 |
| 35+ | 0.1732 | 0.1588 | 0.00491 | 6.28 | 0.103 |
| 36+ | 0.269 | 0.19378 | 0.00729 | 6.29 | 0.12 |
| 37+ | 0.259 | 0.1874 | 0.00225 | 6.27 | 0.08 |
| 38+ | 0.245 | 0.1394 | 3.77E-4 | 6.24 | 0.02 |
| 39+ | 0.1757 | 0.1261 | 2.46E-4 | 6.2 | 0.017 |
| 40+ | 0.1709 | 0.1039 | 2E-4 | 6.18 | 0.025 |
| 41+ | 0.12833 | 0.09669 | 1.8E-4 | 6.17 | 0.03 |
| 42+ | 0.1138 | 0.10626 | 1.79E-4 | 6.13 | 0.02 |

# LIST OF FIGURES

*Figure 1*. Layout showing the configuration of Cryo modules consisting of SC QWRs and solenoids

*Figure 2*. Potential bucket calculated for different charge states for APF period 1 tuned for *q = 34+*

*Figure 3*. Alternating phase variation and energy profile along the APF periods

*Figure 4*. Variation of beam co-ordinates and divergence of 37+ and range spanned by different charge states at end of APF#2 with solenoid field

*Figure 5*. Transmission efficiency of multiple charge state

*Figure 6*. Final distribution of nine charge states from *34+* to *42+* accelerated to 6.2 MeV/u

*Figure 7*. Transmission efficiency with mismatch factor for different charge states. Inset shows the orientation of the mismatched ellipse (red showing the acceptance ellipse)

*Figure 8.* Transmission efficiency (E> 6 MeV/u) and variation of longitudinal and transverse emittance (rms and total) with random transverse misalignments introduced in QWRs and solenoids over a range of ± 100 μm. Blue line shows the corresponding values without any misalignments

# LIST OF FIGURES

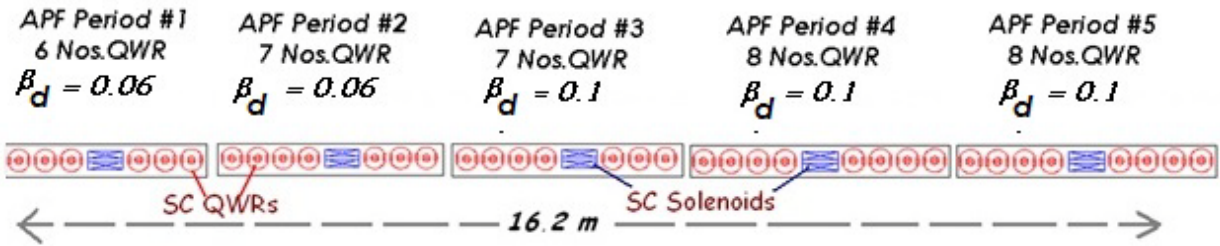

*Figure 1*. Layout showing the configuration of Cryo modules consisting of SC QWRs and solenoids

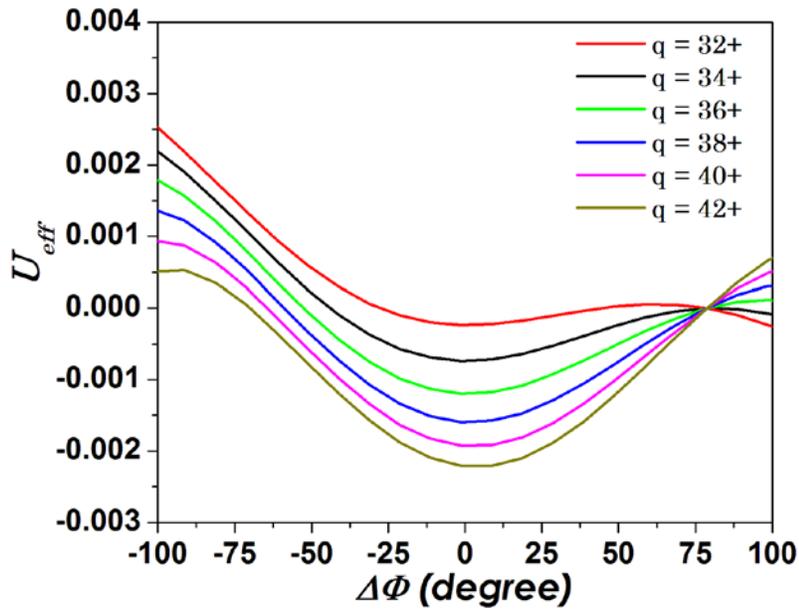

*Figure 2*. Potential bucket calculated for different charge states for APF period 1 tuned for *q = 34+*

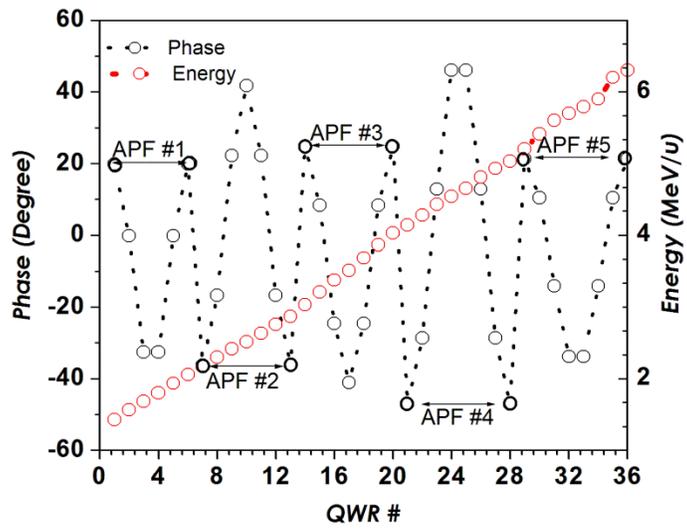

*Figure 3.* Alternating phase variation and energy profile along the APF periods

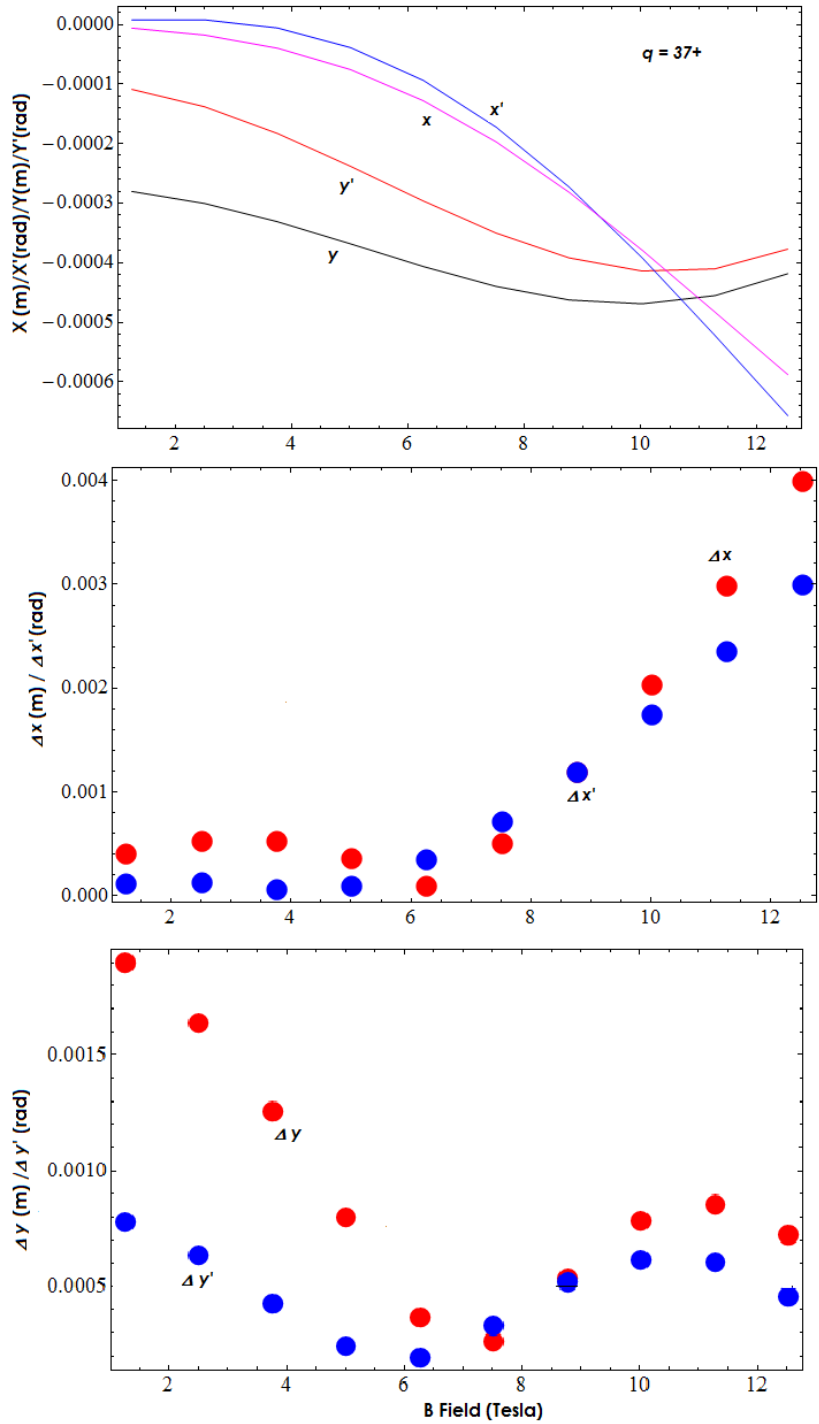

*Figure 4*. Variation of beam co-ordinates and divergence of 37+ and range spanned by different charge states at end of APF#2 with solenoid field

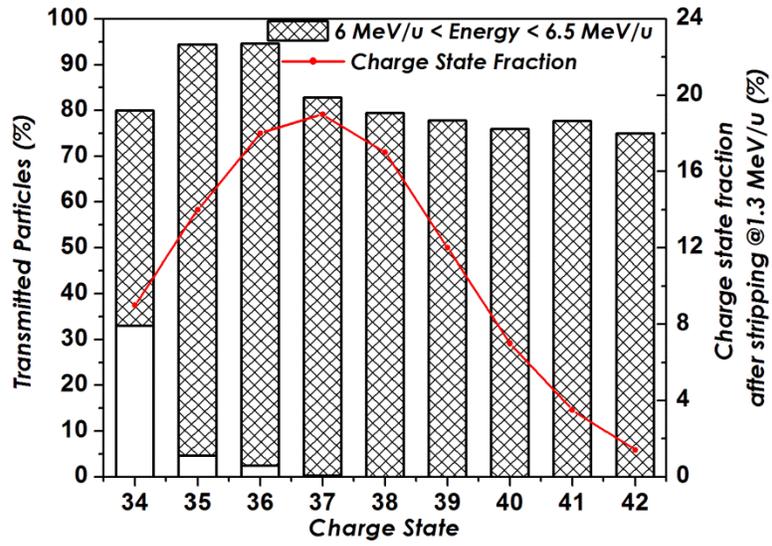

*Figure 5.* Transmission efficiency of multiple charge state

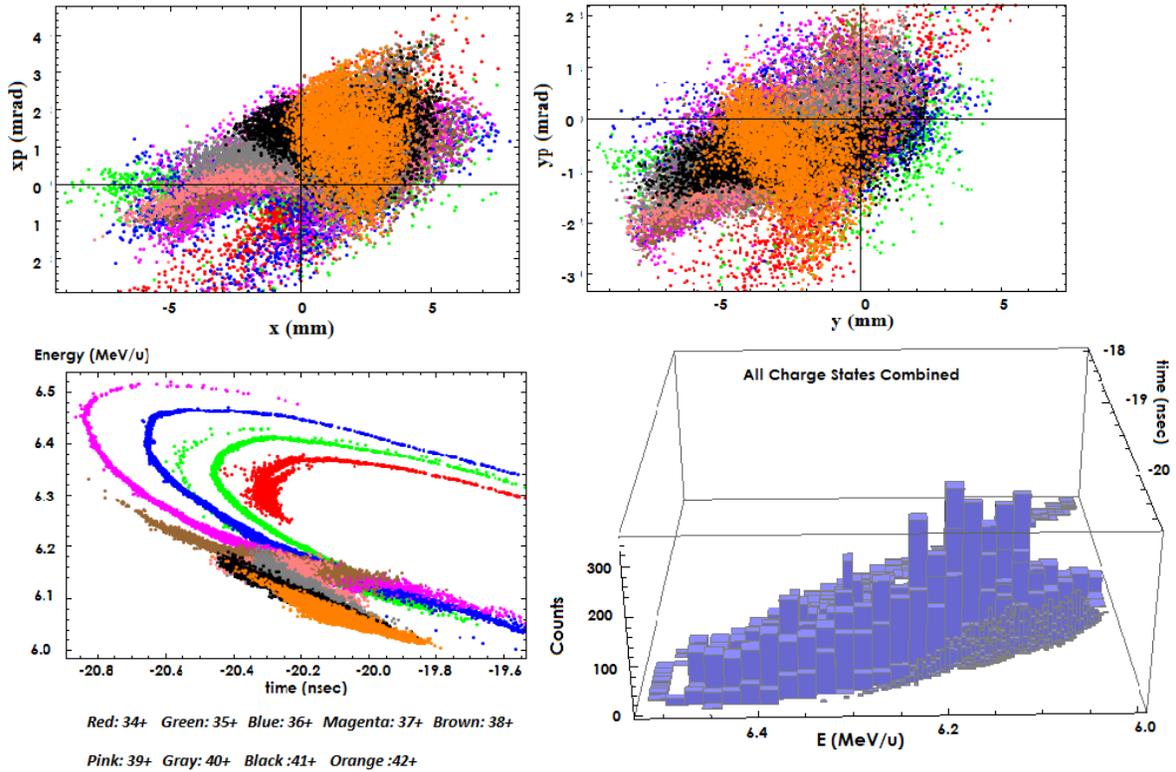

*Figure 6.* Final distribution of nine charge states from *34+* to *42+* accelerated to 6.2 MeV/u

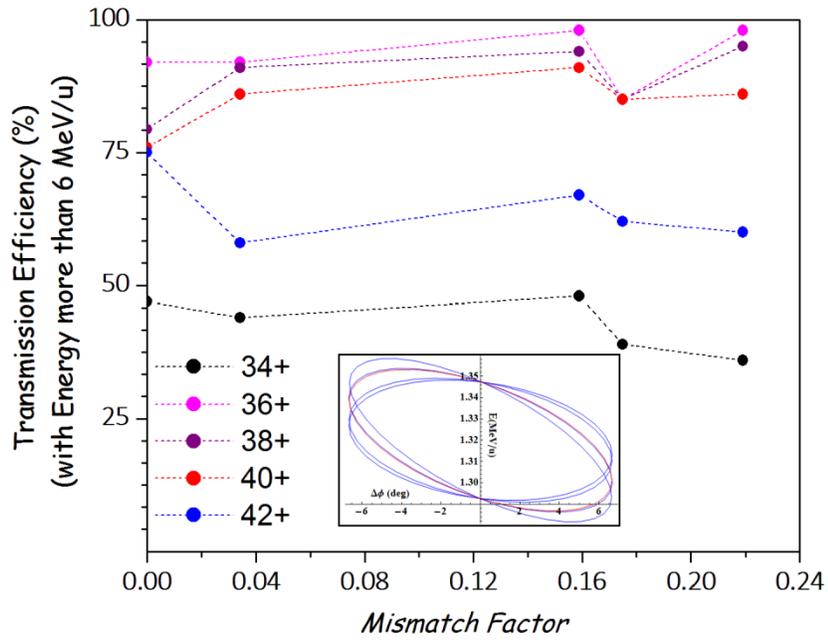

*Figure 7.* Transmission efficiency with mismatch factor for different charge states. Inset shows the orientation of the mismatched ellipse (red showing the acceptance ellipse)

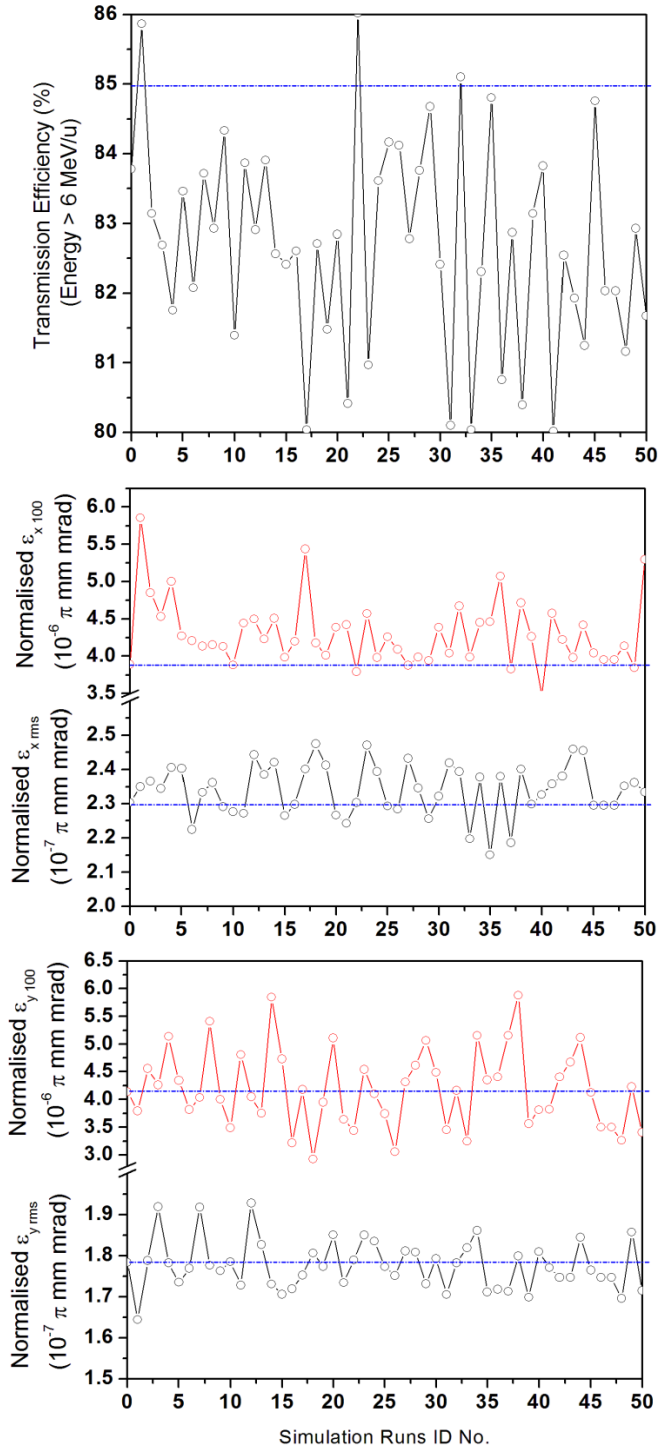

Figure 8. Transmission efficiency (E > 6 MeV/u) and variation of longitudinal and transverse emittance (rms and total) with random transverse misalignments introduced in QWRs and solenoids over a range of ± 100 μm. Blue line shows the corresponding values without any misalignments.